\documentclass{pasj00}

\begin{document}
\SetRunningHead{T.R.Saitoh and J.Koda}{Acceleration Method of Neighbor Search with GRAPE and Morton-ordering.}
\Received{2003/04/14}
\Accepted{2003/05/20}

\title{Acceleration Method of Neighbor Search  \\
with GRAPE and Morton-ordering}

\author{Takayuki R. \textsc{Saitoh}}
\affil{Division of Physics, Graduate School of Science,
 Hokkaido University, N10W8, Sapporo 060-0810, Japan.}
\email{takayuki@astro1.sci.hokudai.ac.jp}

\author{Jin \textsc{Koda}\altaffilmark{1}}
\affil{National Astronomical Observatory, 2-21-1, Osawa, Mitaka, Tokyo, 181-8588, Japan.}
\email{jin.koda@nao.ac.jp}

\altaffiltext{1}{JSPS Research Fellow}


%

\maketitle

\begin{abstract}
We describe a new method to accelerate neighbor searches on GRAPE,
i.e. a special purpose hardware that efficiently calculates gravitational forces
and potentials in $N$-body simulations. In addition to the gravitational
calculations, GRAPE simultaneously constructs the lists of neighbor particles that are
necessary for Smoothed Particle Hydrodynamics (SPH).
However, data transfer of the neighbor lists from GRAPE to the host
computer is time consuming, and can be a bottleneck.
In fact, the data transfer can take about the same time as the calculations
of forces themselves.
Making use of GRAPE's special treatment of neighbor lists,
we can reduce the amount of data transfer if we search neighbors in the
order that the neighbor lists, constructed in a single GRAPE run, overlap
each other.
We find that the Morton-ordering requires very low additional calculation
and programming costs, and results in successful speed-up on data transfer.
We show some benchmark results in the case of GRAPE-5. Typical reduction
in transferred data becomes as much as 90\%.
This method is suitable not only for GRAPE-5, but also GRAPE-3 and
the other versions of GRAPE.
\end{abstract}

\section{Introduction}

GRAPE (GRAvity PipE; Sugimoto et al.\ 1990) is a special purpose 
hardware that calculates Newtonian gravitational forces efficiently 
in large scale $N$-body simulations. 
A series of GRAPE was developed by Ito et al.\ (1990; GRAPE-1),
Ito et al.\ (1991; GRAPE-2),
Okumura et al.\ (1993; GRAPE-3),
Makino et al.\ (1997; GRAPE-4), and Kawai et al.\ (2000; GRAPE-5).
GRAPE is connected with, and controlled by a typical workstation 
or PC. The host computer requests of GRAPE to calculate gravitational 
forces. 
Force integration and particle pushing are all done on the host computer. 
Owing to its high performance, GRAPE has been a powerful tool for 
solving astronomical $N$-body problems, such as those related to
star cluster  evolution \citep{makino96}, black hole spiral-in \citep{makiebi96}, 
formation of planets \citep{kokubo98}, and formation of central 
cusps in dark matter halos \citep{fukushige97}. 

In addition to the efficient gravitational calculations, GRAPE 
performs parallel gathering of neighbor particles, and returns neighbor 
lists to the host computer if requested. 
Since searching neighbors is one of the most time-consuming routines 
in some particle simulations with close interactions, such as 
the Smoothed Particle Hydrodynamics (SPH), the neighbor lists 
from GRAPE are advantageous in speeding up those simulations. 
Thus SPH is often combined with $N$-body calculations using GRAPE. 
Pioneering work on the GRAPE-SPH method was done by
\citet{ume93}, using GRAPE-1A.
\citet{steinmetz96} reported the high performance of the GRAPE-SPH 
method using GRAPE-3. 
The GRAPE-SPH method has been successfully applied to a number of 
topics, e.g., fragmentation of molecular clouds \citep{klessen97}, 
and galaxy formation \citep{steinmetz95, weil98, mori99, koda00a, koda00b}.

Despite the high performance of GRAPE-SPH, 
searching neighbors is still a massive routine in full 
calculations \citep{steinmetz96}. 
In particular, the data transfer of neighbor lists between 
GRAPE and the host computer is a bottleneck for speed-up. 
\citet{steinmetz96} pointed out that, owing to the specification 
of GRAPE, the amount of the data transfer can be reduced 
if the neighbor lists for particles, returned from GRAPE 
at once, overlap each other (see Section 2 in details); 
in the case of GRAPE-3 that can construct 8 neighbor lists
for 8 particles  simultaneously, if the lists are completely the same, 
the communication time for the 8 lists becomes as fast as that
for a single list. In order to make the lists overlap at least 
partially, 
\citet{steinmetz96} sorted the particles into the $X$-coordinate 
order before the GRAPE call, because the particles, having similar 
neighbor lists, must have similar positions, and thus similar 
$X$-coordinates. This approach reduced the time consumed in
entire neighbor searches by 10-20 \%, in simulations using
a few tens of thousands of particles and GRAPE-3.

This approach, however, becomes less effective when the number of 
particles increases, because the particles with 
similar $X$-coordinates are more likely to have very different 
$Y,Z$-coordinates. In this paper, we introduce the other ordering method,
i.e. the Morton ordering method, for GRAPE neighbor searches, which keeps
track of original $3$-dimensional particle coordinates, and are
independent of the number of  particles. We show some test calculations
of neighbor searches using GRAPE-5 and Morton ordering. 
The Morton ordering method has been suggested for some parallel tree 
algorithms for $N$-body simulations \citep{barnes89, warren95}, 
and now, is applied to searching neighbors in GRAPE-SPH.

We briefly review some GRAPE hardware specifications, 
related to searching neighbors in \S 2, and Morton ordering 
in \S 3. Test calculations and results are shown in \S 4 and  
\S 5, respectively. Summary appear in \S 6.

\section{GRAPE: Specification for Neighbor Search}\label{sec:grape}

A GRAPE series is a special purpose board, similar to a graphic board,
used to accelerate gravitational force calculations in $N$-body problems.
It is connected with, and controlled by a host computer, i.e. a
typical workstation or PC. The host computer sends particle positions
and masses to GRAPE, and GRAPE calculates gravitational forces and
potentials, sending them back to the host computer.
GRAPE can also return neighbor particle lists, and thus, is suitable
for simulations with close interactions, such as SPH simulations.
The data transfer of the neighbor lists, however, takes much more
time than that for the others, i.e. for a single particle, typically
sixty words for a neighbor list should be transferred, while only ten
words are transferred for mass, position, force, potential, gravitational
softening, and radius of the particle.	
We will introduce a method to speed up this data transfer
in \S \ref{sec:morton}.
Though we describe the case for the fifth version of the GRAPE series
(GRAPE-5), the method is also well suited for GRAPE-3 and the other
versions of GRAPE. Detailed designs of the GRAPE-5 hardware are
described in \citet{kawai00}. Hence, we give a brief
review, and necessary details for our new neighbor searching method.


The main engine of a GRAPE-5 board is composed of G5 chips.
The G5 chip is a custom LSI chip, which calculates gravitational
forces and potentials. One GRAPE-5 board has eight G5 chips,
each of which calculates the forces on 12 particles
simultaneously, hence one board calculates the forces on
$8 \times 12=96$ particles at once.
Gravitational force ${\bf f}_{i}$ on a particle $i$ is derived
by first calculating the force ${\bf f}_{ij}$ between two particles
($i$ and $j$), and then
summing them up among all particles $\Sigma_j {\bf f}_{ij}$.
The G5 chip can also check whether a particle $j$ is a
neighbor of a particle $i$, by comparing the square of the distance
$r_{ij}^2$ between $i$ and $j$, with the square of the radius $h_i$
of $i$, as $r_{ij}^2<h_i^2$.

The neighbor lists, output from the G5 chips, are stored in special memories
on the GRAPE-5 board. There are two memory units on a single GRAPE-5
board, each of which stores the neighbor lists from four of the eight
G5 chips on a board, and thus, for $4 \times 12=48$ particles.
GRAPE-5 does not keep the neighbor lists in a simple lengthy manner,
such that all neighbors of the 48 particles occupy individual memory
space, but in a way that keeps the lists as particle indices and flags.
For example, when one particle has a neighbor list of
(8,11,22,41,49) and another has (3,7,11,23,41),  these two
lists are kept as a convolved particle index list (3,7,8,11,22,23,41,49)
and binary flags (01,01,10,11,10,01,11,10).
The host computer receives these indices and flags from GRAPE, and
deconvolves these into individual neighbor lists for the two particles.
One neighbor memory unit, of the two on a GRAPE-5 board, stores
the neighbor lists for the 48 particles. Thus the above convolved list
and binary flags are made for the 48 particles.

This particular operation for the neighbor lists provides room to
speed up the data transfer, and hence, the neighbor search.
Considering the case that each particle has $n_s=60$ neighbor
particles, if the 48 particles have completely different neighbors,
the data transferred from GRAPE-5 to the host computer are
$48 \times 60 = 2880$ words for the convolved list, and 2880
words for the binary flags. On the other hand, if we can arrange
the 48 particles so that they have perfectly identical neighbor
lists, the amount of data is significantly reduced to 60 words for the
convolved list and  60 words for the binary flags, 
which means that the communication time is reduced by 1/48
\footnote {
In actual operations, a particle index and binary flag, for a neighbor
particle, are not separately treated. GRAPE stores them in a single
64-bits memory block; the higher 48-bits for the flag, and the lower
16-bits for the index  (see Kawai et al. 2000 for details).}.
Hence, we can reduce the communication time for searching
neighbors with GRAPE, by arranging the 48 particles so that their
neighbors overlap significantly.

\section{GRAPE Neighbor Search with Morton Ordering}\label{sec:morton}

A GRAPE-5 board searches neighbors for 96 particles simultaneously
in a single GRAPE run. In large $N$-body simulations, the GRAPE run
is repeated $N/96$ times. Each of the two memory units on 
a GRAPE-5 board keeps the neighbor lists for $96/2 = 48$
particles in a single run, and the lists are transferred from GRAPE
to the host computer.
According to the GRAPE specifications in \S \ref{sec:grape},
we can reduce the amount of the transferred data if we 
choose the 48 particles, for a single memory in a single run,
so that their neighbors overlap each other.
The cost of the data transfer is reduced by increasing
the fraction of the overlap.

In order to make the neighbor lists overlap,
we should choose 48 intrinsically neighboring particles,
i.e. particles with similar coordinates, for a single GRAPE run.
Based on this idea, \citet{steinmetz96} sorted all the
particles according to their $X$-coordinates, and
succeeded in reducing the communication cost between
GRAPE-3 and a host computer.
This method expects that the particles arranged by
$X$-coordinates would more frequently have similar
$(X,Y,Z)$-coordinates than randomly distributed
particles. However, this method becomes less effective
in very large GRAPE-SPH simulations, because the radius $h$
for searching neighbors becomes smaller in larger simulations,
and thus, two particles with similar $X$-coordinates would
more frequently have quite different $Y,Z$-coordinates,
which makes the separation of the two more than $h$.
Therefore, we suggest the use of Morton ordering,
rather than $X$-coordinate ordering. Morton ordering
naturally translates the $(X,Y,Z)$-coordinates into a 1-D space,
with sufficiently maintaining the original 3-D structure. Morton
ordering has been suggested for a parallel tree code for gravitational
calculations \citep{barnes89}.

\begin{figure}
\begin{center}
\FigureFile(80mm,80mm){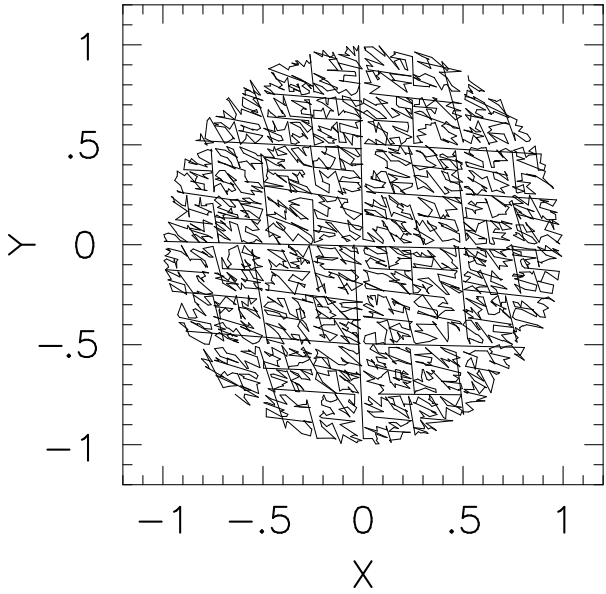}
\caption{Demonstration of the Morton-ordering in a 2-D case for N=3,000.
Randomly distributed particles are connected in a Morton order.
\label{fig1}}
\end{center}
\end{figure}

In the Morton ordering, the 3-D coordinates ($X,Y,Z$) =
($0.x_1 x_2 x_3..., 0.y_1 y_2 y_3..., 0.z_1 z_2 z_3...$) of
a particle is translated into a 1-D {\it key} as $0.x_1 y_1 z_1 x_2 y_2 z_2...$.
Then the particles are sorted according to those keys.
Since those 1-D keys sufficiently have the memory of the original
3-D coordinates, the two particles with similar keys lie close to each
other in the 3-D space as well.
In actual operations,  the key is constructed in a binary space,
and hence, can be simply produced by bit-shift and add.
Thus the costs for the key construction in calculations
and the coding by a programmer are quite low.
Additional time for sorting is also negligible.
A demonstration of the Morton ordering in a 2-D case for 3,000
particles is shown in Figure \ref{fig1}.
The particles are randomly and uniformly distributed in a unit
circle, and connected  in the Morton order (key order)
with the single stroke of a pen.
It is evident that the Morton ordering arranges the particles
in such a way that  those with similar 2-D coordinates lie close to each
other in the key space (1-D),
and that the particles with similar keys must have quite similar
neighbor lists. Hence the Morton ordering is effective to make
the neighbor lists of the 48 particles overlap.

\begin{center}
--------- Figure 1 ---------
\end{center}

\section{Test Calculations}
We test the efficiency of the above new method (GRAPE+Morton ordering),
in comparison with two other methods using GRAPE.
We distribute particles in space, search neighbors for those particles using GRAPE,
and measure the time consumed for the neighbor search.
Before starting the GRAPE neighbor search, we rearrange the particles
(1) in a {\it random} order (hereafter, R-ordering), i.e. with no rearrangement,
(2) in a $X$-coordinate order (X-ordering), and 
(3) in a Morton order (M-ordering).
The last two orderings will actively make the 48 neighbor lists, stored on a single
GRAPE memory unit, overlap each other, which improves the efficiency
of the GRAPE neighbor search as discussed in \S \ref{sec:morton}.

In actual calculations, such as cosmological and galaxy formation simulations,
there appear various density distributions.  Matter is uniformly distributed in the
early stage of the Universe, gradually assembled and collapsed by gravity,
and then, form nearly isothermal objects.
Hence we adopt spherically symmetric
density profiles  $\rho(r) \propto r^n$ with the index $n$
of $0.0$ (uniform) and $-2.0$ (isothermal), for test calculations.
The density profiles are constructed by randomly and uniformly distributing
particles in a unit sphere, and stretching the distribution 
by means of a radial coordinate
transformation, i.e. $r_{new} = r^{3/(3+n)}_{old}$.
We also test the Hernquist profile \citep{hernquist90} as a realistic density model of
dark matter halo, i.e. 
\begin{equation}
	 \rho(r) = \frac{1}{2 \pi} \frac{a}{r} \frac{1}{(r+a)^3}, \label{eq:hernquist}
\end{equation}
where the core and truncation radii
are set to $a = 0.1$ and $r_{\rm max} = 1.0$, respectively.
The number of particles, in the test calculations, is changed from 10,000 to 100,000
every 10,000, which may be possible numbers for actual SPH simulations with the
direct $O(N^2)$ calculations of GRAPE-5.
The neighbor search radius $h$ of each particle is set at the distance of its $n_s$th
nearest particle, and we set $n_s=60$ when no descriptions are given explicitly.
This definition of $h$ is often used in SPH calculations.
 We repeat the neighbor search 10 times for each test calculation, and average them
to get benchmark results, since the results are slightly swayed in individual runs.

For the test calculations, we use a single GRAPE-5 board connected with an Alpha 264
processor computer with a clock frequency of 833MHz, which is one of the GRAPE systems
in the Mitaka Under Vineyard (MUV), run underground at the National Astronomical
Observatory of Japan.

\section{Results}\label{sec:result}

\subsection{Consumption Time}

\begin{figure}
\begin{center}
\FigureFile(80mm,80mm){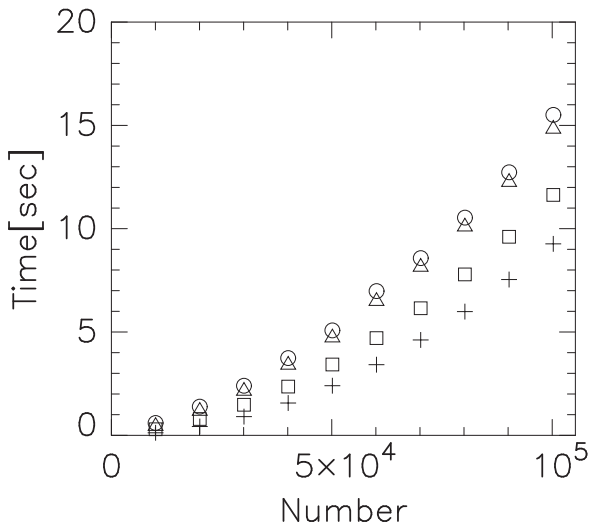}
\caption{Consumption time v.s. number of particles, in the case for
the density profile index of $n = -2$.
Circles are times for R-ordering, triangles are for X-ordering,
and squares are for M-ordering. Crosses indicate the time consumed
for GRAPE calculations without data transfer, which cannot intrinsically
be suppressed in the above three methods.
\label{fig:time-2}}
\end{center}
\end{figure}

Table \ref{tab:time} summarizes the consumption times in neighbor searches with
GRAPE, in cases using random ordering (R-ordering), $X$-coordinate ordering 
(X-ordering), and Morton ordering (M-ordering). The tabulated times include both
the data transfers between GRAPE and the host computer, and the calculations
for searching neighbors in GRAPE. The above three orderings differ only in their
data transfer times.
Figure \ref{fig:time-2} shows a corresponding plot for the isothermal density profile
($n=-2$), where
the times for the GRAPE calculations without data transfer are also drawn as
crosses. The differences between crosses and the other marks indicate
the times for data transfer and its overhead.

\begin{center}
---------  Table 1 ---------
\end{center}

Generally, the consumption times show no clear difference among three density
profiles,
because GRAPE intrinsically does $O(N^2)$-operations, which do not depend on
density profiles. Hence we hereafter discuss only the case for the density profile
of an index $n=-2$.
It is evident that M- and X-ordering work faster than R-ordering for any $N$, and
that M-ordering is more efficient than X-ordering.

In our GRAPE system, M-ordering works twice as fast as R-ordering for
$N=10,000$, while X-ordering does only 1.3 times as fast.
M-ordering is 1.5 times faster than R-ordering for $N=50,000$, while X-ordering
is 1.1 times faster.
Both  M- and X-ordering apparently become less effective for larger $N$ on
the basis of total calculation time (Figure \ref{fig:time-2}), while M-ordering keeps
its efficiency even in larger $N$ (see \S \ref{sec:frac}).
This is because both orderings save only the communication costs, i. e. $O(N)$-operations,
between GRAPE and the host computer. However calculations
in GRAPE, i.e. $O(N^2)$, become more dominant for larger $N$.

\begin{center}
---------  Figure 2 ---------
\end{center}

For $N=100,000$, the largest number in our tests, X-ordering becomes inefficient,
i.e. consuming almost the same amount of time as R-ordering, while M-ordering is still
1.4 times faster than R-ordering.
Therefore, M-ordering is best suited for neighbor searches with GRAPE.

\subsection{Data Compression Factor}\label{sec:frac}
In \S \ref{sec:grape} we described how the communication time between GRAPE and
a host computer is reduced if we make the neighbor lists for 48 particles, kept in
a memory unit in a single GRAPE run, overlap each other. In order to describe
how much the lists overlap, we define the mean data compression factor of neighbor lists as 
\begin{equation}
f = \frac{N_b^{trans}}{N_b^{total}}, \label{definef}
\end{equation}
where $N_b^{total}$ is a simple sum of the numbers of neighbors for all $n_p$ particles,
and  $N_b^{trans}$ is the number of neighbors actually transferred from GRAPE
to the host computer, according to the GRAPE specifications (\S \ref{sec:grape}).
We note that this factor does not depend on the speeds of a host computer and
an interface between GRAPE and the host computer. The communication time is reduced
in proportion to $f$.

If we consider a single GRAPE run and the case that each of the $n_p=48$
particles has 60 neighbors, $N_b^{total}$ is $48 \times 60 = 2880$.
If the neighbors of the 48 particles are completely independent,
$N_b^{trans}$ becomes $48 \times 60 = 2880$. Then the compression factor
becomes $f=1$, meaning no compression.
If the neighbor lists are perfectly identical, $N_b^{trans}$ becomes $60$ as
described in \S \ref{sec:grape}, and then the compression factor takes
its theoretical minimum, i.e. $f=1/n_p$ [Note that this is an insubstantially
ideal case (see A.\ref{sec:festim})]. For test calculations with a large
number of particles, the GRAPE run must be repeated many times. 
Then we average $f$ in all the
data transfer for neighbor lists in all the runs.

Table \ref{tab:fraction} lists $f$ for all the test calculations, and a corresponding
plot for the isothermal profile ($n=-2$) are presented in Figure \ref{fig:fraction-2}.
The compression factors $f$ of R- and X-orderings increase with the number of particles $N$.
The $f$ for X-ordering is efficiently as small as 0.5 for $N=10,000$; however it increases
to about 0.8 for $N=100,000$. R-ordering shows almost no data compression ($f=0.98$),
that is, 98\% are left for data transfer in the case of $N=100,000$.
Hence X- and R-orderings do not work well for large $N$ calculations.
On the other hand, M-ordering keeps $f$ almost constant at the low value of 0.13
for all the $N$s (Figure \ref{fig:fraction-2}).
This is why M-ordering is still effective in large $N$ calculations.
The low value of $f=0.13$ means that the neighbor lists, sorted simultaneously
on a single GRAPE memory unit, overlap almost perfectly (87\%),
and thus, implies that there is little room for further improvement.
Therefore we conclude that our new method (GRAPE+Morton ordering)
is the best for neighbor searches using GRAPE.

\begin{center}
---------  Table 2 ---------
\end{center}

\begin{center}
---------  Figure 3 ---------
\end{center}

\subsection{Dependence on $n_s$}
Figure \ref{fig:nstime} shows the $n_s$ dependence of consumption time in the case of
the isothermal density profile ($n=-2$) and $N=100,000$.
We tested the range of $n_s=30-120$, which is used in actual SPH calculations.
Basically, the consumption times increase with $n_s$, because the number of neighbor
particles, transferred from GRAPE to the host, increases with $n_s$.
We note, however, that there is another effect that suppresses the increase of time.
The larger $n_s$ means larger radii (volumes) of particles, and thus, indicates larger
overlap of their neighbor lists. This reduces data transfer, and results in saving time.
Figure \ref{fig:nsfraction} shows this effect; the overlap fractions $f$ decrease
with increasing $n_s$. In our realistic range of $n_s$, M-ordering shows the best
overlap fraction, and thus is the best in any $n_s$.

\begin{center}
---------  Figure 4 ---------
\end{center}
\begin{center}
---------  Figure 5 ---------
\end{center}

\section{Summary}

We have reviewed the specifications of a special purpose hardware
called GRAPE, and
 introduced a new method which can speed up neighbor searches in large
particle simulations using GRAPE. The main conclusions are the following:


1.
We introduced a new method, that is, arranging particles in a Morton order
before performing GRAPE calculations. This method saves the communication cost
between GRAPE and its host computer. The cost for additional programming
is very low.


2. 
We compare this Morton-ordering method with some previous methods,
and conclude that the Morton-ordering method is much more effective.
In a case where the total particle number is $N=10,000$, the Morton-ordering
method is twice as fast as a simple neighbor search with GRAPE in our GRAPE system.


3.
Communication between GRAPE and its host computer can be minimized
if the neighbor lists, stored in a single GRAPE memory unit,
overlap each other. 
The Morton-ordering method reduces the communication by about 90\%,
thus leaving little room for further improvement.


4.
The Morton-ordering method becomes less effective for larger particle
simulations, as do the other previous methods, because 
$O(N^2)$-calculations, other than data transfer, become dominant.
However, it is still efficient for
simulations with $N=100,000$.


5.
The communication increases with the typical number of neighbor particles.
The Morton-ordering method is the best in any number that is usually used
in SPH calculations.


6.
We showed the efficiency of the Morton-ordering method only for GRAPE-5.
However, it is also suitable for the other versions of GRAPE. In fact, this method has been
effectively used for galaxy formation simulations using GRAPE-3
\citep{koda00a, koda00b}.\\

T.R.S. would like to thank Asao Habe  for his encouragement, Takashi Okamoto
and Tamon Suwa for many useful discussions about coding.
We thank Jun Makino, the referee, for useful suggestions, which
improved Appendix.
T.R.S. was supported by the Sasakawa Scientific Research  Grant from
The Japan Science Society(14-096). J.K. was financially supported by
the Japan Society for the Promotion of Science for Young Scientists.
Numerical computations were carried out on the GRAPE system
(project ID:g02a09) at the Astronomical Data Analysis Center of 
the National Astronomical Observatory, Japan, 
which is an inter-university research institute of astronomy 
operated by the Ministry of Education, Science, Culture, and Sports.


\clearpage

\appendix

\section{Performance Estimation}\label{sec:pfestim}
The total calculation time for neighbor search with GRAPE
will be modeled as
\begin{equation}
T = T_{\rm h} + T_{\rm g} + T_{\rm t},\label{eq:ttime}
\end{equation}
where $T_{\rm h}$, $T_{\rm g}$ and $T_{\rm t}$ stand for
the time consumed on the host computer, that on GRAPE, and
that on data transfer of neighbor lists from GRAPE to the host computer,
respectively.
$T_{\rm h}$ and $T_{\rm g}$ are modeled as
\begin{eqnarray}
T_{\rm h} &=& c_{\rm h} N \\
T_{\rm g} &=& c_{\rm g} N^2
\end{eqnarray}
where $c_{\rm h}$ and $c_{\rm g}$ represent the miscellaneous
calculation time per particle on the host computer, and the time
spent on a two body interaction on GRAPE, respectively.
$N$ is the number of particles in calculation.

The total number of neighbor particles transferred from GRAPE to
the host computer is $N n_s f$, where $n_s$ is the typical number of
neighbors for one particle, and $f$ is the data compression factor
defined in \S \ref{sec:frac}.
$T_{\rm t}$ would take a form as
\begin{equation}
T_{\rm t} = c_{\rm t} N n_s f,\label{eq:tmtrans}
\end{equation}
where $c_{\rm t}$ is the time spent on data transfer per neighbor particle.
For our GRAPE system (GRAPE-5 and a host computer with an Alpha 264
processor 833MHz),
we obtain the coefficients by fitting, and list them on Table \ref{tab:fitting}.
Figure \ref{fig:met} shows a plot of estimated v.s. measured $T$ for
R-, X-, and M-orderings. Different symbols are used for different orderings.
All the points are well on the proportional line (solid).

\begin{center}
---------  Table \ref{tab:fitting} ---------
\end{center}
\begin{center}
---------  Figure \ref{fig:met} ---------
\end{center}

\section{Theoretical Estimate of $f$}\label{sec:festim}
The data compression factor $f$ takes the minimum of $1/n_p$
in an {\it insubstantially} ideal case that all the $n_p$ particles have an identical
neighbor list, however, its actual minimum, occurred in calculation, would be larger.
We estimate the $f$ in a {\it thoughtfully} ideal case, and
compare it with the results of M-ordering.
We consider the case that the $n_p$ particles (see \S\ref{sec:grape})
are selected very successfully, i.e. the case that their neighbor lists overlap
almost as much as possible. Since we have shown that $f$ does not depend on the
distribution of particles (\S \ref{sec:result}), we assume that $N$ particles are
distributed in a unit sphere with a uniform density, i.e. $\rho=3 N/4\pi$.
In the ideal case, the $n_p$ particles
themselves must be closest neighbors each other, and be clustered in a small region.
In the following we assume that this region has a spherical form.

\begin{figure}
\begin{center}
\FigureFile(80mm,80mm){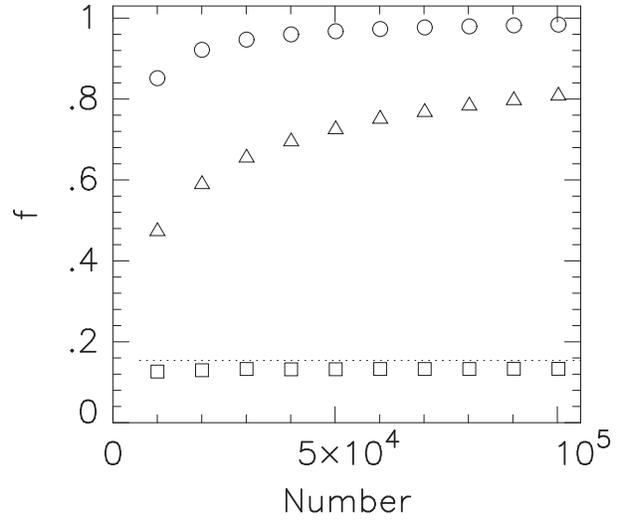}
\caption{Data compression factor v.s. number of particles, in the case for
the density profile index of $n = -2$. The same marks are used as
Figure \ref{fig:time-2}.
Dotted line indicates a theoretical estimate of $f$ in the case that
the $n_p=48$ particles are spherically distributed (A.\ref{sec:festim}).
 \label{fig:fraction-2}}
\end{center}
\end{figure}

\begin{figure}
\begin{center}
\FigureFile(80mm,80mm){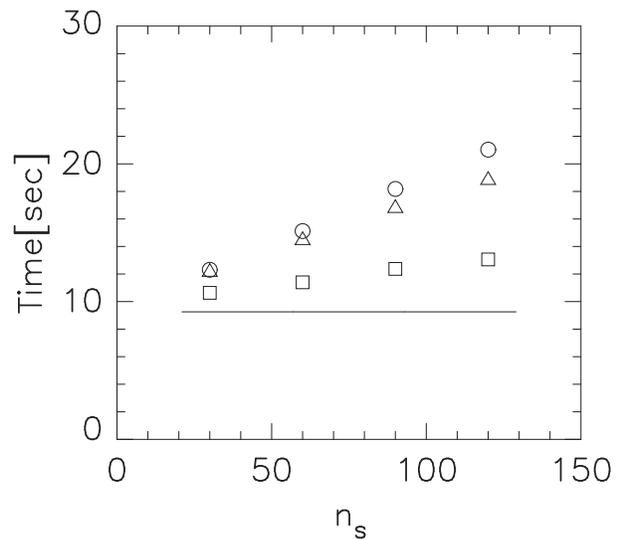}
\caption{
$n_s$ dependence of consumption time for $n = -2.0$
and $N=100,000$.
Circles are times for R-ordering,
triangles are for X-ordering, squares are for M-ordering.
Solid line indicates the consumption time for GRAPE
calculations without data transfer.
\label{fig:nstime}}
\end{center}
\end{figure}

The $n_p$ particles are distributed in a sphere with the radius
$r_p = (n_p/N)^{1/3}$. If we take into account that some of the $n_p$ particles
are on the surface of the sphere, and that each of them has $n_s$ neighbors
and a radius $r_s = (n_s/N)^{1/3}$, then all the neighbors of $n_p$
particles will be {\it at least} within a sphere of the radius $r_a=r_p+r_s$.
Hence the number of neighbor particles, stored in a GRAPE memory and
transferred from GRAPE to the host computer, becomes
\begin{equation}
N^{trans}_b = ( n_p^{1/3} + n_s^{1/3} )^3.\label{eq:ntsphere}
\end{equation}
Since the total accumulated number of neighbors for the $n_p$ particles is
$N^{total}_b = n_p n_s$, $f$ is calculated as
\begin{equation}
f = \frac{ ( n_p^{1/3} + n_s^{1/3} )^3 }{ n_p n_s}.\label{eq:festim}
\end{equation}
This $f$ does not depend on $N$, and approaches $1/n_p$ when $n_s \rightarrow \infty$.
Figure \ref{fig:fraction-2}  shows this $f$ (dotted line).
The results of M-ordering (squares) are close to the estimated $f$ of this
ideal case.

\begin{figure}
\begin{center}
\FigureFile(80mm,80mm){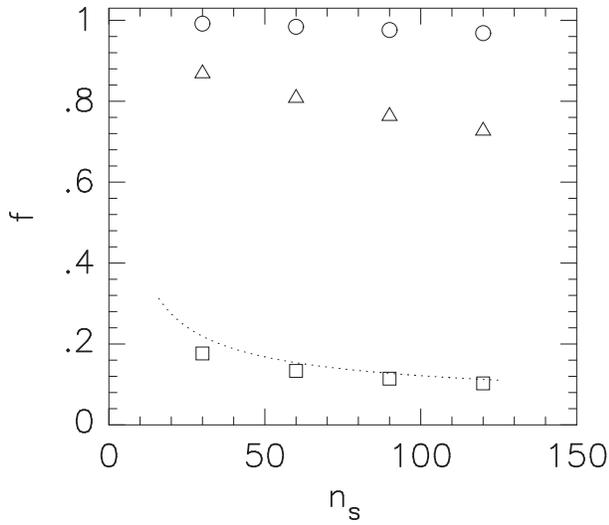}
\caption{
$n_s$ dependence of data compression factor $f$ for $n = -2.0$.
The same marks are used as Figure \ref{fig:nstime}.
Dotted line indicates a theoretical estimation of $f$ in the case
that the $n_p=48$ particles are spherically distributed (see A.\ref{sec:festim})
\label{fig:nsfraction}}
\end{center}
\end{figure}

\begin{figure}
\begin{center}
\FigureFile(80mm,80mm){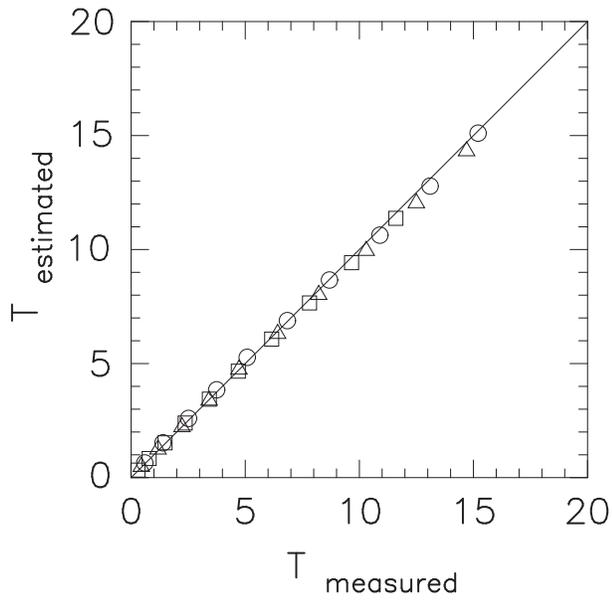}
\caption{
Estimated v.s. measured times consumed on neighbor search.
The same marks are used as Figure \ref{fig:nstime}.
Solid line is a proportional line. Estimated time is calculated by
eq. (\ref{eq:ttime}).}
\label{fig:met}
\end{center}
\end{figure}

Note, we here assumed that all the neighbors are closely packed in
an even spherical space with a radius $r_a=r_p+r_s$. However,
this assumption is valid only in an infinite limit of $n_s$, because the actual
space occupied by small $n_s$ particles must have uneven surface,
which is completely enclosed by our assumed sphere.
Then $N^{trans}$, and $f$, is smaller than that estimated
by eq.(\ref{eq:ntsphere}). Hence in Figure \ref{fig:fraction-2},
M-ordering gives slightly smaller $f$ than the estimated one for the spherical case.
This difference becomes smaller with increasing $n_s$, which is confirmed in
Figure \ref{fig:nsfraction} (dotted line and squares).

Eq. (\ref{eq:festim}) gives a thoughtful minimum of $f$ that can be occurred in
actual calculations. The fair coincidence of this minimum value with those from
M-ordering gives us the confidence that M-ordering is the ideal method for neighbor
search in GRAPE-SPH.


\clearpage

\begin{table}
\caption{Time consumed by neighbor search}\label{tab:time}
\begin{center}
\begin{tabular}{rrrrrrrrrrrr}
\hline \hline 
 & \multicolumn{3}{c}{Power index $n = 0.0$} & & \multicolumn{3}{c}{Power index $n = -2.0$} & &
 \multicolumn{3}{c}{Hernquist model}\\
\cline{2-4} \cline{6-8} \cline{10-12} 
Number & R-ord. & X-ord. & M-ord. & & R-ord. & X-ord. & M-ord. & & R-ord. & X-ord. & M-ord. \\
\hline
10000 &  0.61 &  0.45 &  0.30 && 0.60 & 0.45 & 0.30 && 0.55 & 0.40 & 0.28 \\
20000 &  1.40 &  1.18 &  0.79 && 1.40 & 1.18 & 0.79 && 1.34 & 1.11 & 0.75 \\
30000 &  2.41 &  2.17 &  1.48 && 2.52 & 2.23 & 1.49 && 2.36 & 2.09 & 1.42 \\
40000 &  3.75 &  3.44 &  2.36 && 3.75 & 3.42 & 2.37 && 3.56 & 3.24 & 2.29 \\
50000 &  5.07 &  4.75 &  3.43 && 5.09 & 4.73 & 3.43 && 4.95 & 4.58 & 3.33 \\
60000 &  6.95 &  6.52 &  4.69 && 6.85 & 6.42 & 4.70 && 6.64 & 6.23 & 4.60 \\
70000 &  8.89 &  8.38 &  6.17 && 8.69 & 8.22 & 6.16 && 8.49 & 8.23 & 6.05 \\
80000 &  10.9 &  10.3 &  7.80 && 10.9 & 10.3 & 7.82 && 10.3 & 9.89 & 7.63 \\
90000 &  13.1 &  12.5 &  9.62 && 13.1 & 12.5 & 9.66 && 12.5 & 12.0 & 9.42 \\
100000 &  15.2 &  14.7 &  11.6 && 15.2 & 14.7 & 11.6 && 15.1 & 14.5 & 11.4 \\
\hline
\end{tabular}
\end{center}
Test calculations for a unit sphere with a density profile of
$\rho \propto r^n$, and the Hernquist profile. Consumption time is presented in units of seconds.
Particles are randomly distributed, and R-, X-, and M-ordering rearrange
the particles in {\it random} order, in $X$-coordinate order,
and in Morton order, respectively, before GRAPE calculations.
\end{table}

\begin{table}
\caption{Data Compression Rate $f$.}\label{tab:fraction}
\begin{center}
\begin{tabular}{rrrrrrrrrrrr}
\hline \hline
  & \multicolumn{3}{c}{Power index $n = 0.0$} & & \multicolumn{3}{c}{Power index $n = -2.0$} & &
 \multicolumn{3}{c}{Hernquist model}\\
\cline{2-4} \cline{6-8} \cline{10-12} 
Number & R-ord. & X-ord. & M-ord. & & R-ord. & X-ord. & M-ord. & & R-ord. & X-ord. & M-ord. \\
\hline
  10000 & 0.85 & 0.48 & 0.12 && 0.85 & 0.47 & 0.13 && 0.85 & 0.47 & 0.13\\
  20000 & 0.92 & 0.60 & 0.13 && 0.92 & 0.59 & 0.13 && 0.93 & 0.59 & 0.14\\
  30000 & 0.95 & 0.67 & 0.13 && 0.95 & 0.66 & 0.13 && 0.95 & 0.65 & 0.14\\
  40000 & 0.96 & 0.72 & 0.13 && 0.96 & 0.70 & 0.13 && 0.96 & 0.70 & 0.14\\
  50000 & 0.97 & 0.75 & 0.13 && 0.97 & 0.73 & 0.13 && 0.97 & 0.73 & 0.14\\
  60000 & 0.97 & 0.77 & 0.13 && 0.97 & 0.76 & 0.13 && 0.97 & 0.76 & 0.14\\
  70000 & 0.98 & 0.78 & 0.13 && 0.98 & 0.77 & 0.13 && 0.98 & 0.77 & 0.14\\
  80000 & 0.98 & 0.79 & 0.13 && 0.98 & 0.78 & 0.13 && 0.98 & 0.78 & 0.14\\
  90000 & 0.98 & 0.81 & 0.13 && 0.98 & 0.79 & 0.13 && 0.98 & 0.79 & 0.14\\
100000 & 0.98 & 0.82 & 0.13 && 0.98 & 0.80 & 0.13 && 0.98 & 0.80 & 0.14\\
\hline
\end{tabular}
\end{center}
\end{table}

\begin{table}
\caption{Timing constants for performance estimation}\label{tab:fitting}
\begin{center}
\begin{tabular}{cc}
\hline
Parameter & Constant \\
                     & (sec) \\
\hline
$c_{\rm h}$ &  $1.8 \times 10^{-5}$ \\
$c_{\rm g}$ & $9.0 \times 10^{-10}$  \\
$c_{\rm t}$ &  $7.3 \times 10^{-7}$ \\
\hline
\end{tabular}
\end{center}
\end{table}
\clearpage

\end{document}